\documentclass[prd,superscriptaddress,twocolumn,showpacs,%
amsmath,amssymb]{revtex4-1}

\usepackage{slashed}
\usepackage{graphicx}
\usepackage[breaklinks,colorlinks=true]{hyperref}

\newcommand{\tr}{\text{tr}}
\newcommand{\bk}{\boldsymbol{k}}
\newcommand{\calG}{\mathcal{G}}
\newcommand{\bx}{\boldsymbol{x}}

\newcommand{\bomega}{\boldsymbol{\omega}}
\newcommand{\bSigma}{\boldsymbol{\Sigma}}

\usepackage{color}

\begin{document}
\title{Chiral vortical effect with finite rotation, temperature, and curvature}

\author{Antonino Flachi}
\affiliation{Department of Physics,
  and Research and Education Center for Natural Sciences,
  Keio University, 4-1-1 Hiyoshi, Yokohama, Kanagawa 223-8521, Japan}
  
\author{Kenji Fukushima}
\affiliation{Department of Physics, The University of Tokyo,
  7-3-1 Hongo, Bunkyo-ku, Tokyo 113-0033, Japan}

\begin{abstract}
  We perform an explicit calculation of the axial current at finite
  rotation and temperature in curved space.  We find that finite
  curvature and mass corrections to the chiral vortical effect satisfy
  a relation of the chiral gap effect, that is, a fermion mass-shift
  by a scalar curvature.  We also point out that a product term of the
  angular velocity and the scalar curvature shares the same
  coefficient as the mixed gravitational chiral anomaly.  We discuss
  possible applications of the curvature induced chiral vortical
  effect to rotating astrophysical compact objects described by the
  Kerr metric.  Instead of direct calculation we assume that the
  Chern-Simons current can approximate the physical axial current.  We
  make a proposal that the chiral vortical current from rotating
  compact objects could provide a novel microscopic mechanism behind
  the generation of collimated jets.
\end{abstract}
\maketitle

\section{Introduction}

The chiral vortical effect (CVE) refers to the topological axial
current induced by rotation of chiral matter.  An analytical formula
for the current has been originally derived microscopically for a
Dirac matter distribution in a rotating frame~\cite{Vilenkin:1978hb}
and applied to neutrino fluxes from rotating black
holes~\cite{Vilenkin:1979ui}.  More detailed calculations for general
field theories were later reported in Ref.~\cite{Vilenkin:1980zv}.
Interest in the CVE has been reignited by an analogous topological
phenomenon called the chiral magnetic effect (CME) that refers to the
generation of an electric current due to the axial anomaly in the
presence of an external magnetic field (see
Ref.~\cite{Kharzeev:2013jha} and contributions therein).  From the
analogy with the CME, we can naturally anticipate that angular
momentum would induce a similar effect, i.e., a chiral vortical
current along the rotation axis.  With the rotation axis chosen along
the $z$ direction, the chiral vortical current can be written, to
linear order in the angular velocity $\omega$ and for massless
fermions, as (see Ref.~\cite{Kharzeev:2015znc} for a recent review):
\begin{equation}
  j_{R/L}^z = \pm\biggl(\frac{T^2}{12}
  + \frac{\mu_{R/L}^2}{4\pi^2} \biggr)\omega \;,
\label{eq:CVE}
\end{equation}
with $R/L$ indicating the right-handed and left-handed sectors
separately.  Anomalous hydrodynamics~\cite{Son:2009tf}, AdS/CFT
correspondence~\cite{Erdmenger:2008rm}, chiral kinetic
theory~\cite{Stephanov:2012ki} have all substantiated
Eq.~\eqref{eq:CVE}.

In Ref.~\cite{Landsteiner:2011cp} a conjecture relating the
current~\eqref{eq:CVE} to the anomalies with gauge and gravitational
fields has been proposed (see Ref.~\cite{Basar:2013qia} for related
discussions).  Based on the Kubo formula for the chiral vortical
conductivity with metric perturbations in the framework of fluid
dynamics~\cite{Landsteiner:2011iq}, it was shown that the coefficients
of the chemical potential $\mu_{R/L}^2$ and of the temperature $T^2$
in Eq.~\eqref{eq:CVE} are respectively proportional to the chiral
anomalies in the gauge and the gravitational sectors.  It is not
entirely clear whether this conjecture is true in general and it is
even possible to have a $T^2$ part of the CVE even when there is no
perturbative anomaly (see Refs.~\cite{Golkar:2012kb}).  However, the
suggestion is intriguing as the non-vanishing transport coefficients
appear as a manifestation of anomalies.

Generally it is not obvious whether the CVE is rooted in the mixed
gravitational chiral anomaly or not, as the relation between the two
may be quite indirect, as we will stress in the present work.
So far, the established fact is that the coefficients appearing in the
expression of the anomaly and in the chiral vortical current are
common for some reason but this fact does not necessarily require that
one can be derived from the other.

A way to clarify the situation would be to explicitly calculate
gauge-invariant physical observables, namely, the axial current
expectation value at finite rotation (with the angular velocity
$\omega$), temperature $T$, and curvature (with the scalar curvature
$R$), not resorting to the anomaly.  Here we emphasize that a
curved-space setup on top of the ordinary CVE at finite $T$ and
$\omega$ provides between the CVE and the gravitational chiral
anomaly, though, we do not insist on any precise relation between the
CVE and the anomaly.  Anticipating the step-by-step derivations of
Sec.~\ref{sec2}, we shall write our final result down below:
\begin{equation}
  j_{R/L}^z = \pm \biggl(\frac{T^2}{12} - \frac{m^2}{8\pi^2}
  - \frac{R}{96\pi^2} \biggr)\,\omega\,,
\label{eq:CVE_mod}
\end{equation}
where $m$ is the fermion mass.  For simplicity we dropped finite
chemical potential terms $\propto \mu_{R/L}^2$, but it is not
difficult to recover them.  It is the last term $\propto R$ in
Eq.~\eqref{eq:CVE_mod} that would hint an indirect mechanism for the
same coefficient as the chiral anomaly.

Beyond the formal significance in clarifying how thermal and
geometrical effects mirror into each other, and how this impacts on
the CVE formula, recently, it is of increasing interest to investigate
the physics of relativistic rotating matter in heavy-ion collision
experiments.  Although the theoretical description of a spinning fluid
is not yet fully understood~\cite{Becattini:2009wh} (see
Ref.~\cite{Becattini:2015tpl} for recent discussions related to the
present work), microscopic field-theoretical calculations are
feasible.  Rotating quark matter possibly created in heavy-ion
collisions may accommodate a non-trivial phase diagram as described in
Refs.~\cite{Jiang:2016wvv,Chernodub:2016kxh}.  In heavy-ion
collisions, moreover, not only thermal effects but also those of
strong magnetic fields play a critical role~\cite{McLerran:2013hla}.
Then, as emphasized in Ref.~\cite{Chen:2015hfc}, an effective chemical
potential associated with rotation would topologically induce a
non-zero density~\cite{Hattori:2016njk}, which is one concrete
manifestation of the chiral pumping effect~\cite{Ebihara:2015aca}.
(As pointed out in Ref.~\cite{Ebihara:2016fwa} the partition function
obtained in Ref.~\cite{Chen:2015hfc} encompassed such an induced
density.)  Because hot and dense matter created in heavy-ion
collisions is rapidly expanding, a finite curvature associated with
three dimensionally expanding geometries is expected to modify the
above-mentioned estimates according to Eq.~\eqref{eq:CVE_mod}.

Another intriguing corollary of the above arguments is associated to
the physics of astrophysical jets, whose formation mechanism (i.e.,
acceleration and collimation) is surrounded by many open questions.
It is certainly an attractive idea to draw a connection between the
CVE and the microscopic nature of jets from compact astrophysical
sources, as discussed in the present work.

\section{Explicit Calculation}
\label{sec2}

Our goal here is to compute the expectation value of the axial current
in curved space directly using the propagator in a rotating system,
i.e.,
\begin{equation}
  j_A^\mu(x)
  = -i \lim_{x'\to x} \tr\bigl[\gamma^\mu \gamma_5 S(x,x')\bigr]\;.
  \label{eq:jA}
\end{equation}
Thus, all we need is the explicit form of the propagator $S(x,x')$ at
finite $T$ on the background of a rotating curved geometry.  To
construct the propagator, it is intuitively clearer to treat
rotational and geometrical features separately.

We employ Riemann normal coordinates $\xi$ around a point $x$
(identified by $\xi=0$) and consider the coincident limit of the
fermion propagator.  Using Riemann normal coordinates significantly
simplifies the analysis since the Christoffel symbols at $x$ are all
vanishing and the Dirac matrices are just those in flat spacetime.  A
finite rotation $\omega$ can then be introduced as a small
perturbation.

The coincident limit of the propagator in normal coordinates for
$\omega=0$ takes the form,
\begin{equation}
  S_0(x,x'\to x) = \int\frac{d^4 k}{(2\pi)^4}\,(-\gamma^\mu k_\mu+m)\,
  \calG(k)\;.
\label{eq:prop}
\end{equation}
Here $k$ is a momentum conjugate to $\xi$ and $\calG(k)$ is a known
function involving metric derivatives~\cite{parker} as
\begin{align}
  \calG(k) &= \Biggl[ 1 - \biggl(A_1+iA_{1\alpha}
    \frac{\partial}{\partial k_\alpha} - A_{1\alpha\beta}
      \frac{\partial}{\partial k_\alpha \partial k_\beta}
      \biggr) \frac{\partial}{\partial m^2} \notag\\
  &\qquad + A_2\biggl(\frac{\partial}{\partial m^2}\biggr)^2 \Biggr]
      \frac{1}{k^2-m^2} + \cdots\;,
\label{eq:G}
\end{align}
where $A_1$ represents a coefficient with mass-dimension 2 that is
expressed in terms of Riemann tensors at $x$. Analogously,
$A_{1\alpha}$ is a mass-dimension 3 coefficient, and
$A_{1\alpha\beta}$, $A_2$ are mass-dimension 4 coefficients, involving
spin operators.  Explicit expressions for these coefficients can be
found in Ref.~\cite{parker}.  It is easy to argue by dimensional
analysis that higher-order terms represented by the ellipses are
suppressed at sufficiently high $T$, as we will explicitly see later.

For technical simplicity, in what follows we require two conditions to
be satisfied.  The first is that of stationarity, i.e., all metric
components are time independent and the temporal components of the
metric are space independent.  We require this to utilize the standard
Matsubara formalism valid for systems in thermal equilibrium.  This
condition may be relaxed at the price of using the more complicated
real-time formalism to include thermal effects.  We note that rotation
induces a space-dependence in $g_{00}$ at $\omega^2$ order, but for
our purposes it is necessary to go only to linear order in $\omega$.
This is not a particularly restrictive assumption, since, for small
$\omega$, we can always reduce the metric to a form compatible with
this assumption by means of a conformal transformation.

The second condition we require is that all metric components are $z$
independent and the $z$ components of the metric are space
independent.  This condition corresponds to choosing the rotation axis
along the $z$ direction, and removes the $z$ dependence of the spin
operators of the rotation generators.

Thanks to the simplicity of Riemann normal coordinates, the
calculations are straightforward.  In this setup, the temporal
direction is not distorted, and thus the propagator is a function of
$t-t'$.  This allows us to define an energy conjugate to $t-t'$ that
is nothing but $k_0$ in Eq.~\eqref{eq:prop}.  By applying the rotation
generator, we can write the rotating propagator with $\bomega$ as
\begin{equation}
  S(\bx,\bx',k_0) = e^{\bomega\cdot \frac{1}{2}\bSigma
    \frac{\partial}{\partial k_0} }
  S_0(\bx,\bx',k_0)\;,
\end{equation}
for $\bx'\sim \bx$.  (Note that $\bx$ is the center of rotation, so
that there is no orbital term.)  Here, $\bSigma$ is a spin operator
defined by $\Sigma^i = \epsilon^{ijk}\frac{i}{4}[\gamma_j,\gamma_k]$.
Using our assumption of $\omega$ being small, we can proceed to expand
in powers of $\omega$.  To 0th order, we can replace $S(\bx,\bx',k_0)$
with $S_0(\bx,\bx',k_0)$.  Then, using the symmetry properties of the
Riemann tensors, we can readily convince ourselves that
$j_A^\mu\bigr|_{\omega=0}=0$.  This is expected: 
even in curved space the axial current is vanishing as long as there
is no rotation.

To 1st order in $\omega$, the spin operator produces a difference
leading to a non-zero Dirac trace.  So, the whole quantity is
proportional to
$\tr[\gamma_5\gamma^\mu\gamma^{\mu'}\gamma^{\nu'}\gamma^\nu]
=4i\epsilon^{\mu\mu'\nu'\nu}$, a trait common to
anomaly calculations.  Some algebra gives
\begin{equation}
  j_A^\mu = i\,\epsilon^{\mu\mu'\nu'\nu}
  \omega_{\mu'\nu'}\,\int\frac{d^4 k}{(2\pi)^4}\,
  \frac{\partial}{\partial k_0} \,k_\nu\, \calG(k)\;,
\end{equation}
where we used a two-index representation of the angular velocity as
$\omega^i = \epsilon^{ijk}\omega_{jk}$.  Summation over the Matsubara
frequencies $k_0$ is understood after the $k_0$ derivative is taken in
the integrand.  Using Eq.~\eqref{eq:G}, we see that the first term
returns the well-known formula of the CVE{}.  That is, defining the
energy dispersion $\varepsilon_k=\sqrt{\bk^2+m^2}$, the CVE arises
from
\begin{equation}
  \begin{split}
  & \int\frac{d^4 k}{(2\pi)^4}\frac{\partial}{\partial k_0}
  \frac{k_0}{k^2-m^2} = -i\int\frac{d^3 k}{(2\pi)^3}\,
  n_F'(\varepsilon_k) \\
  &= \frac{i}{\pi^2}\int_0^\infty dk\,\biggl(\varepsilon_k
  -\frac{m^2}{2\varepsilon_k}\biggr)\, n_F(\varepsilon_k)\;,
  \end{split}
\label{eq:finite_m}
\end{equation}
with $n_F(z)$ being the Dirac-Fermi distribution function.  The
integral above amounts to
$\frac{i \Gamma(2)\zeta(2)}{2\pi^2}T^2 = \frac{i}{12}T^2$ in the
$m\to 0$ limit, from which we correctly arrive at Eq.~\eqref{eq:CVE}.

The most interesting correction to the axial current emerges from the
second term in Eq.~\eqref{eq:G}.  From {textbook}~\cite{parker}
$A_1=R/12$, with the momentum integration being almost the same as
the previous one apart from the mass derivative, we have:
\begin{equation}
  \frac{\partial}{\partial m^2}\int\frac{d^4 k}{(2\pi)^4}\,
  \frac{\partial}{\partial k_0}\,\frac{k_0}{k^2-m^2}
  = -\frac{i}{2}\int\frac{d^3 k}{(2\pi)^3}\, 
  \frac{n_F''(\varepsilon_k)}{\varepsilon_k}\;.
\end{equation}
In the $m\to 0$ limit the above integral yields $-i/(8\pi^2)$.  Therefore,
together with the first term, the total current turns out to be
\begin{equation}
  j_A^z = \biggl( \frac{T^2}{12} - \frac{m^2}{8\pi^2}
  - \frac{R}{96\pi^2} + \cdots \biggr)\,\omega\;,
\end{equation}
in which neglected ellipses are higher order terms such as $R\,m^2/T^2$
for small $m$ and $R$.  This proves our central result of
Eq.~\eqref{eq:CVE_mod}. In the formula above, from
Eq.~\eqref{eq:finite_m}, we inferred to lowest order the finite-$m$
corrections: $iT^2/12 \to i(T^2/12 - m^2/8\pi^2)$. We note that, as we
shall discuss shortly, at zero temperature and zero curvature, at
linear order there should be no CVE and the above expression assumes
$m^2\ll T^2$.  Then, according to the chiral gap
effect~\cite{Flachi:2014jra}, a finite scalar curvature shifts the
fermionic mass gap as $m^2 \to m^2 + R/12$, which perfectly explains
the ratio between the second and the third terms in
Eq.~\eqref{eq:CVE_mod}.  (See also Ref.~\cite{Jensen:2012kj} where
the same term $\propto R\omega$ was obtained in a yet different way.)

It is also interesting to point out that the coefficient $1/12$
obtained, for instance in
Refs.~\cite{Landsteiner:2011cp,Golkar:2012kb}, is derived as the
coefficient multiplying the scalar curvature term in the (heat-kernel)
coefficient in Ref.~\cite{Flachi:2014jra}. This number, $1/12$, is
independent of the background geometry.

We can continue the expansion to include higher-order corrections from
Eq.~\eqref{eq:G}.  The next contribution leading to finite corrections
seems to be $A_{1\alpha\beta}$.  This term involves one more mass
derivative,
\begin{equation}
  \frac{\partial}{\partial m^2}\int\frac{d^3 k}{(2\pi)^3}\,
  \biggl[\frac{n_F''(\varepsilon_k)}{\varepsilon_k}
  + \frac{n_F'''(\varepsilon_k)}{3} \biggr] \;\to\;
    -\frac{7\zeta(3)}{16\pi^4 T^2}\;,
\end{equation}
in the $m\to 0$ limit.  It is non-trivial that the above combination
of the integrals is infrared finite, though each has singularity.
This adds a correction to the current by
$\delta j_A^\mu = 3\bar{A}_{100} \cdot 7\zeta(3) / (16\pi^4 T^2)$,
where $\bar{A}_{1\alpha\beta}$ represents a part of $A_{1\alpha\beta}$
without spin operator ~\cite{parker}. However, n the present treatment
with only static deformations, $\bar{A}_{100}$ is zero.  Thus, the
first non-zero correction appears from the second derivative in terms
of $m^2$, that is,
\begin{equation}
  \delta j_A^z = \bar{A}_2\cdot
  \frac{7\zeta(3)}{32\pi^4 T^2}
\end{equation}
with $\bar{A}_2$ being a mass-dimension 4 coefficient given by
$\bar{A}_2=\frac{1}{120}{R_{;\mu}}^\mu+\frac{1}{288}R^2
-\frac{1}{180}R_{\mu\nu}R^{\mu\nu}+\frac{1}{180}
R_{\mu\nu\sigma\tau}R^{\mu\nu\sigma\tau}$.  We stop here and will not
include this correction in our considerations below.

Let us turn to intriguing features of the expansion.  Explaining how a
$T$-independent correction $\propto R$ in Eq.~\eqref{eq:CVE_mod}
appears from finite-$T$ calculations requires a delicate interchange
of the two limits; $m \rightarrow 0$ and $T \rightarrow 0$.  In fact,
if we keep a finite $m$ and take the $T \rightarrow 0$ limit first,
then we would have
$\int\frac{d^3 k}{(2\pi)^3}\, \varepsilon_k^{-1} n_F''(\varepsilon_k)\to 0$,
and no such term $\propto R$ survives, as we already noted.
Therefore, the order of two limits, $T\to 0$ and $m\to 0$ is
important.  Here, we always assume the $m\rightarrow 0$ limit first
and then vary $T$, as the value of $m$ defines the theory, while $T$
is a control parameter that we can adjust externally in physical
situations.

One more comment is due.  The above-mentioned calculations would be
reminiscent of the high-$T$ expansion, but we emphasize that there is
a crucial difference.  If one performs the high-$T$ expansion for the
pressure $p$ for example, the leading term is proportional to $T^4$,
the next leading term $m^2 T^2$, and the further next term $m^4$.  The
important point is that such $m^4$ term in the high-$T$ expansion is
accompanied by a logarithmic singularity, $\ln(m/\pi T)$, which blows
up for both $m\to 0$ and $T\to 0$.  Unlike this, in the present case,
such terms involving $\ln(m/\pi T)$ exactly cancel out, which can be
also confirmed in the heat-kernel expansion.  We remark that a
logarithmic singularity with $R$ should vanish in odd spatial
dimensions.  These interesting observations might be related to the
non-renormalization of the anomaly.

\section{Approximate Estimate of the Axial Current}

For a more general geometry, beyond our simplifying assumptions, even
up to linear order in $\omega$, the direct calculation of the axial
current is impossibly difficult.  We here propose to utilize as a
proxy of the axial current the Chern-Simons current.  As is well
recognized, the Chern-Simons current is not gauge invariant, but we
have empirically known that it can approximate the physical current,
which is the case, for example, for the orbital component of the
photon angular momentum.  The CME is a well known example of the
Chern-Simons current acquiring a physical significance thanks to an
external chemical potential.

The chiral anomaly with gravitational background fields
reads~\cite{Kimura:1969}:
\begin{equation}
  \nabla_\mu j_A^\mu = \frac{1}{384\pi^2}\epsilon^{\mu\nu\rho\lambda}
  R_{\mu\nu}^{~~~\alpha\beta} R_{\rho\lambda\alpha\beta}\,.
\end{equation}
The right-hand side takes the form of a total divergence, from which
the Chern-Simons current can be derived.  In this way we can find the
Chern-Simons current $j_{\rm CS}^\mu$ associated with the
gravitational chiral anomaly as
\begin{equation}
  j_{\rm CS}^\mu = \frac{1}{96\pi^2}\epsilon^{\mu\nu\rho\lambda}
  {\Gamma^\alpha}_{\nu\beta}\Bigl( \partial_\rho {\Gamma^\beta}_{\alpha\lambda}
  +\frac{2}{3}{\Gamma^\beta}_{\rho\sigma}{\Gamma^\sigma}_{\alpha\lambda}
  \Bigr)\;.
  \label{eq:CS}
\end{equation}
Under a coordinate transformation from $x^\mu$ to $x'^\mu$ with
rotation, ${\Gamma^k}_{ij}$ acquire a correction by
$(\partial x'^k/\partial x^r)(\partial^2 x^r)
/(\partial x'^i \partial x'^j)$, that gives not only multiplicative
transform but also additive shift as
\begin{equation}
  \delta{\Gamma^x}_{0y} = -\delta{\Gamma^y}_{0x} = \omega\;.
\end{equation}
Then, up to linear order in $\omega$, the Chern-Simons current takes
the following form:
\begin{equation}
  j_{\rm CS}^\mu = \frac{\omega}{48\pi^2} \bigl(
         {R^0}_{x0x} + {R^0}_{y0y} - {R^x}_{yxy} - {R^y}_{xyx} \bigr)\;.
\end{equation}
The important observation here is that, once the $\omega$ dependence
is extracted, the remaining part is written in terms of the Riemann
tensors only.  In our case, with flat $0$ and $z$ directions, only
${R^x}_{yxy}$ and ${R^y}_{xyx}$ survive, leading to the same
expression as our direct calculation of Sec.~\ref{sec2},
$j_A^z=-\omega R/(96\pi^2)$.  This is a consistency check for our
Ansatz of using the Chern-Simons current as a proxy of the directly
calculated axial current.

The relation between the microscopically computed current and the
Chern-Simons current should be understood in the same way as for the
CME current.  In the CME case, the axial current along the $z$ axis is
proportional to $\epsilon^{z0ij}A_0 \partial_i A_j$, which itself is
gauge variant.  However, once the chemical potential $\mu$ is
turned on, $A_0$ is replaced with $\mu$, and then the rest part is the
field strength tensor and thus gauge invariant;
$\langle j_A^z\rangle \propto \mu B$~\cite{Fukushima:2008xe}.  In this
way, the Chern-Simons current can be interpreted as a physical current
due to the external environment.  Our explicit calculations support
the idea that the derivation of the CME based on the Chern-Simons
current may hold also for the CVE involving  the metric background
with the following correspondence ($\mu$ in the CME) $\leftrightarrow$
($\omega$ in the CVE).

At a glance one may feel not easy to upgrade the Chern-Simons current
to a physical quantity.  In the CME case the subtle point is how $\mu$
can be physical, while $A_0$ is not.  The answer is that $\mu$ is a
holonomy:  $A_0$ itself can be gauged away but $\mu$ is a remainder
that cannot be gauged away under the periodic boundary condition in the
imaginary-time formalism.  In fact it is well known that the Polyakov
loop is a gauge invariant holonomy in non-Abelian gauge theories, and
Abelian imaginary $\mu$ can be defined similarly in a gauge invariant
way.  For our problem with finite $\omega$, it can be gauged away by
rotating coordinate transformation.  We can, however, introduce an
indelible $\omega$ imposing periodic boundary condition.  This
observation is consistent with the results in
Ref.~\cite{Hattori:2016njk,Ebihara:2016fwa} where it is argued that
rotation alone cannot induce any physical consequence.

\section{Astrophysical Jets from the Chiral Vortical Effect}

\begin{figure}
  \includegraphics[width=0.6\columnwidth]{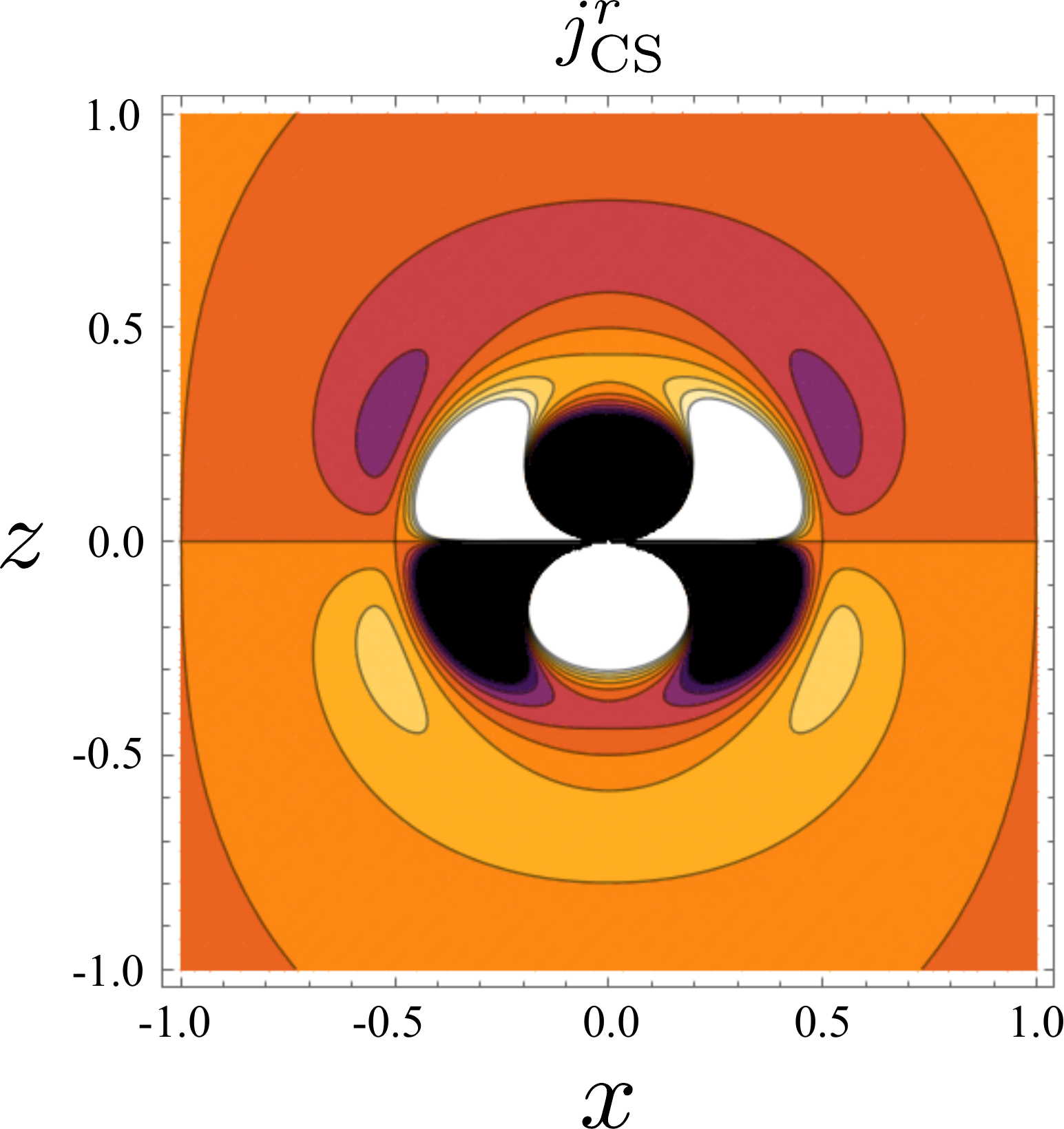} \vspace{1em}\\
  \includegraphics[width=0.6\columnwidth]{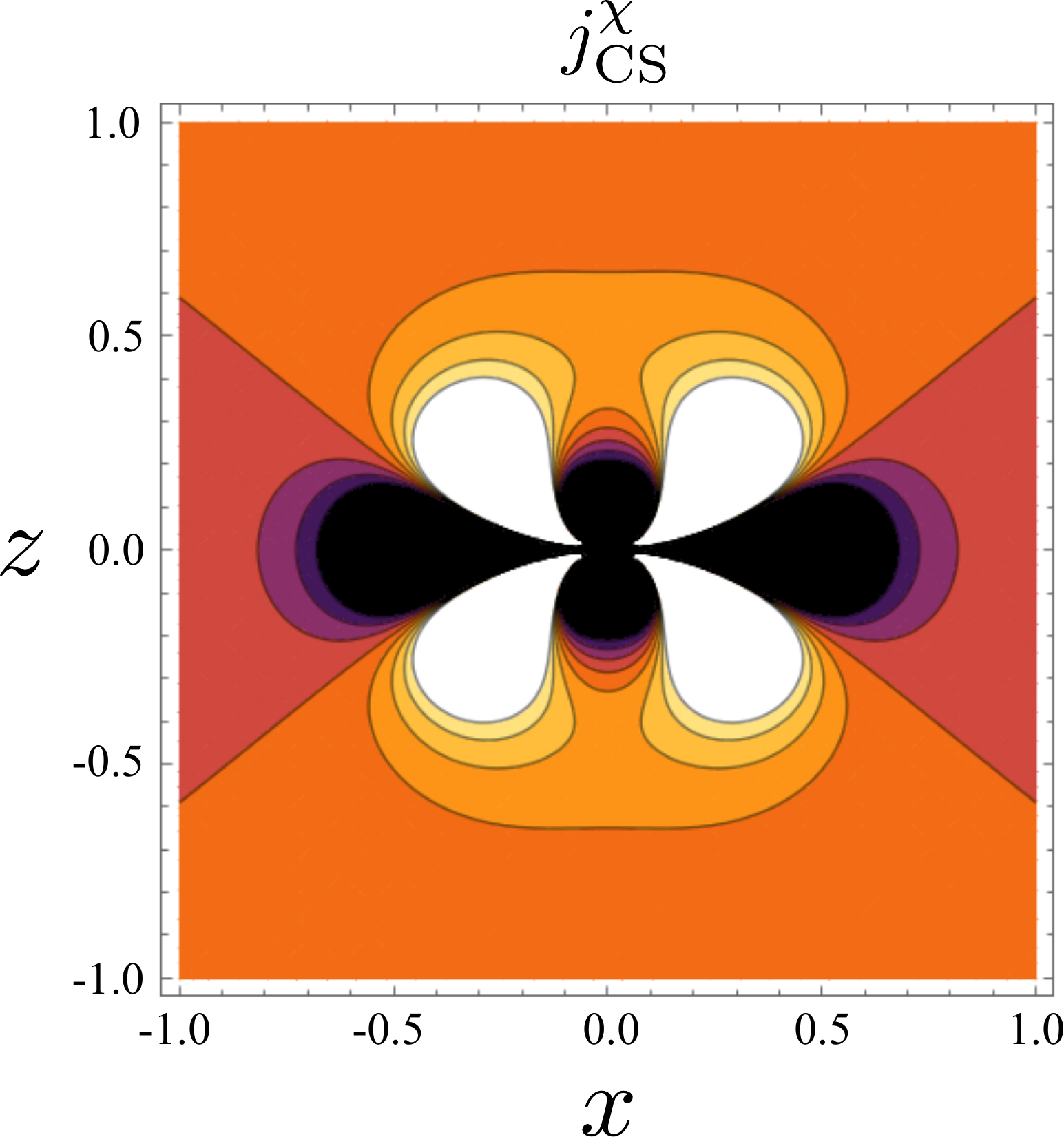}
  \caption{Axial current in the extremal limit in units of $\omega$.
    Light color represents positively large values and dark color
    represents negatively large values.}
  \label{fig:currents}
\end{figure}

Once the above Ansatz of the Chern-Simons current as a proxy of the
physical axial current is accepted, we have a powerful method to
proceed to numerical computations.  Let us consider a rotating
gravitational background, described by the Kerr metric.  It would be a
complicated calculation to evaluate the propagator on the Kerr
geometry, but it is rather straightforward to write the Chern-Simons
current down.  In Boyer-Lindquist coordinates ($t$, $r$,
$\chi=\cos\theta$, $\phi$), after some calculations, we find
$j_{\rm CS}^r\neq0$ and $j_{\rm CS}^\chi\neq0$, while
$j_{\rm CS}^0=j_{\rm CS}^\phi=0$.  Here, instead of showing the full
expressions, let us discuss $j_{\rm CS}^r$ and $j_{\rm CS}^\chi$ in
particular limits only.  For small $\omega$, the current to linear
order in $\omega$ reads,
\begin{equation}
  j_{\rm CS}^r = \frac{3\pi(-3\pi+8rT_B)\chi}{24576 r^6 T_B^4}\omega\;,\quad
  j_{\rm CS}^\chi = \frac{3\pi(-1+3\chi^2)}{6144 r^6 T_B^3}\omega\;.
\label{eq:CSomega}
\end{equation}
where $T_B$ is the black hole temperature (and not the thermodynamic
temperature).  If the thermodynamic temperature is involved, as
discussed in Ref.~\cite{Jensen:2012kj}, spatial derivatives of the
temperature would appear.  In confronting the above expressions with
the formula~\eqref{eq:CVE}, we should remark that in
Eqs.~\eqref{eq:CSomega} both $T_B$ and $r$ are dimensionful
quantities.

The angular dependence in the above results,
$j_{\rm CS}^r\propto \chi=\cos\theta$, indicates the presence of a
current aligned with the rotation axis.  Coming back to the
discussions in Ref.~\cite{Vilenkin:1979ui}, we can associate this
axial current with neutrino flux.  It is then tempting to interpret
the present results in terms of a novel (sharing some similarities
with the Penrose process~\cite{Penrose}) microscopic mechanism for the
generation of collimated astrophysical jets observed in rotating
compact stellar objects (see Ref.~\cite{Meier:2003bi}).
Interestingly, it may be worth noticing that this mechanism would be
generic to all rotating compact objects, and not limited to black
holes.

\begin{figure}
  \includegraphics[width=0.8\columnwidth]{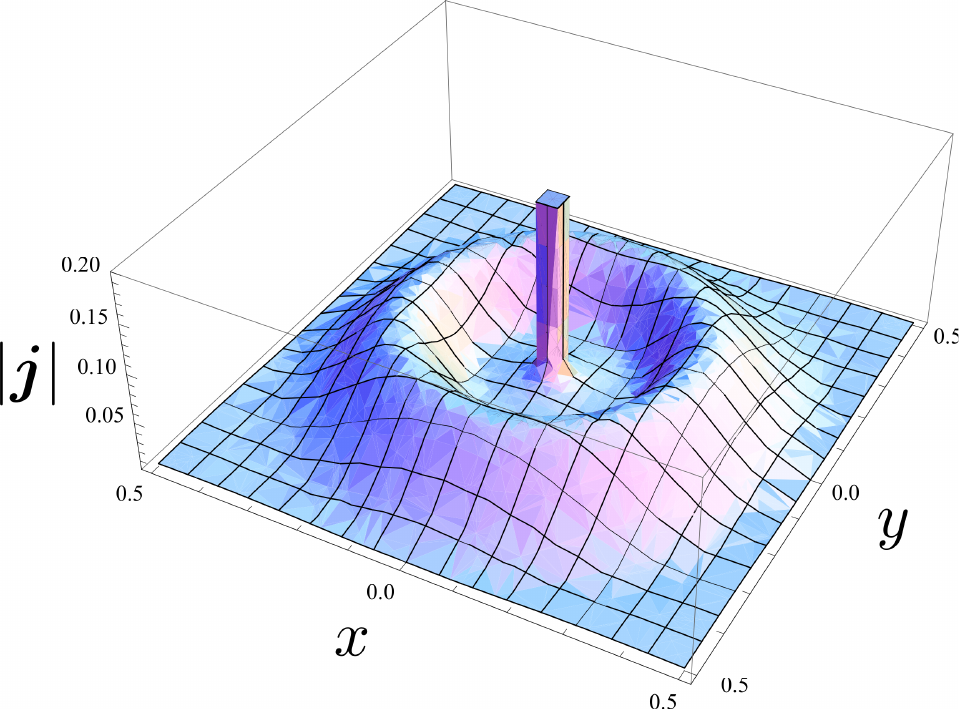}
  \caption{Axial current magnitude in the extremal limit at $z=0.2$ in
    units of $\omega$.}
  \label{fig:current_mag}
\end{figure}

For astrophysical applications it is relevant to examine the extremal
limit $T_B=0$.  One may think, using expressions~\eqref{eq:CSomega},
that the limit $T_B\to 0$ would be singular.  However, before the
expansion in $\omega$, the limit smoothly exists.  It should be noted
that Eq.~\eqref{eq:CSomega} gives the leading order term in an
expansion in powers of $\omega/T_B$. Therefore, we cannot extrapolate
Eq.~\eqref{eq:CSomega} naively to $T_B\to 0$.  The correct result for
the leading $\omega$ order in the extremal limit is
\begin{align}
  j_{\rm CS}^r &= -\frac{(1-2\xi)[\chi^4+4\chi^2 \xi
      (3-8\xi)-48\xi^3(1-\xi)]\chi}{3\pi^2 (\chi^2+4\xi^2)^5}\omega^3 ,
  \label{eq:jr}\\
  j_{\rm CS}^\chi &= \frac{[\chi^6-\chi^4(3+56\xi^2)
      +72\chi^2\xi^2(1+2\xi^2)-48\xi^4]}{3\pi^2 (\chi^2+4\xi^2)^5}\omega^4 ,
  \label{eq:jchi}
\end{align}
where $\xi=r\omega$.  For clarity, we should remark that although the
expansion for small $T_B$ and that for small $\omega/T_B$ do not
commute, the limiting procedure is straightforward and does not pose
any difficulty.  It is interesting that these currents in the extremal
limit become increasingly large for $\chi\to0$ if $\omega$ (and $\xi$)
is small enough.  This feature is very different from
Eq.~\eqref{eq:CSomega}.  The currents in Eqs.~\eqref{eq:jr} and
\eqref{eq:jchi} are plotted in Fig.~\ref{fig:currents}, where we use
the unit in terms of $\omega$ and we set $y=0$ without loss of
generality due to the axial symmetry.  It is not easy to imagine how
the current is spatially distributed from Fig.~\ref{fig:currents}, so
the magnitude of the current is plotted in Fig.~\ref{fig:current_mag}
which shows a 3D jet profile.  As illustrated in these figures, the
currents are strongly peaked near $z\sim 0$ or $\theta\sim \pi/2$.
It is worth noting that heavy and slowly rotating objects generally
exhibits such singular structures, implying that common compact
stellar objects in the universe should be accompanied by a axial
currents as displayed in Fig.~\ref{fig:currents}.  This result implies
that the CVE currents may be a source for the surrounding disk as well
as the astrophysical jets.

We note that Eqs.~\eqref{eq:jr} and \eqref{eq:jchi} are rapidly
damping as $|\boldsymbol{j}|\propto r^{-5}$ at large distance.  This
is so because there is no given chiral charge and no net production of
$j_{\rm CS}^0$ in this case.  In other words the Kerr metric has
$R=0$, so that the leading-order CVE term $\propto \omega R$ is
vanishing.  For a more qualitative estimate, we must
assume a ``freezeout'' radius beyond which free particles are emitted
out.  In this work we will not go further into attempts to quantify
our estimates using the Kerr metric.  In reality black holes could be
charged, and combinations with electromagnetic fields produce more
contributing terms.  Here, we point out a qualitative possibility and
leave quantitative discussions including missing terms for the
future.

It is an intriguing problem to discuss the physical implications of
these currents for hot and dense quark matter in heavy-ion
collisions as well as in astrophysics.  In the same way as to
interpret the chiral anomaly as parity-odd particle
production~\cite{Fukushima:2015tza}, we can give a physical picture
for these currents as extra contributions to phenomena similar to
Hawking radiation (see Ref.~\cite{Vilenkin:1979ui} for discussions
along these lines).  Another interesting application includes
anomalous neutrino transport in rapidly rotating system of black hole
or neutron star mergers (see Ref.~\cite{Yamamoto:2015gzz} for an idea
of anomalous neutrino transport in supernovae and
Ref.~\cite{Gorbar:2016klv} for applications to the early universe).

\section{Conclusions}

In this work we have calculated the axial current expectation value in
curved space at finite temperature.  The chiral vortical effect
receives a correction proportional to the scalar curvature, $R$, which
is consistent with the finite mass correction and the chiral gap
effect.  We point out that such a topologically induced current
$\propto \omega R$ with $\omega$ being the angular velocity has the
same overall coefficient as the Chern-Simons current.  Our argument
parallels that in the derivation of the chiral magnetic effect that is
fully explained by the replacement of $A_0$ with the chemical
potential $\mu$ in the Chern-Simons current.  This physical
augmentation of the Chern-Simons current due to the external
environment offers an interesting theoretical device to approximate
the particle production in non-trivial background geometries.
We have adopted this Ansatz to use the Chern-Simons current as a proxy
of the physical current to the case of a rotating astrophysical body
and have argued that the chiral vortical current may provide a novel
universal microscopic mechanism behind the generation of collimated
jets from rotating astrophysical compact sources.

\begin{acknowledgments}
  We thank Yoshimasa~Hidaka, Karl~Landsteiner, Pablo~Morales, and
  Shi~Pu for discussions.
  K.~F.\ thanks Francesco~Becattini and Kristan~Jensen for comments.
  K.~F.\ was partially supported by JSPS KAKENHI Grants
  No.\ 15H03652, 15K13479, and 18H01211.
  A.~F.\ acknowledges the support of the MEXT-Supported Program
  for the Strategic Research Foundation at Private Universities
  `Topological Science' (Grant No.\ S1511006).
\end{acknowledgments}

\end{document}